\documentclass[conference]{IEEEtran}
\IEEEoverridecommandlockouts
\def\BibTeX{{\rm B\kern-.05em{\sc i\kern-.025em b}\kern-.08em
    T\kern-.1667em\lower.7ex\hbox{E}\kern-.125emX}}

\ifCLASSINFOpdf

\else

\fi

\usepackage{graphicx,changepage}
\usepackage{algorithm,algpseudocode}
\usepackage{amsmath}
\usepackage{verbatim}
\usepackage{mathtools}
\usepackage{amssymb}
\usepackage[mathscr]{euscript}
\usepackage{caption}
\usepackage[numbers]{natbib}
\usepackage[bookmarks=false]{hyperref}
 \pdfoutput=1
\begin{document}

\title{Safe Driving Capacity of Autonomous Vehicles}

\author{\IEEEauthorblockN{ Yuan-Ying Wang}
\IEEEauthorblockA{\textit{Department of Electrical Engineering} \\
\textit{National Taiwan University}\\
Taipei, Taiwan \\
r06921051@ntu.edu.tw}
\and
\IEEEauthorblockN{ Hung-Yu Wei}
\IEEEauthorblockA{\textit{Department of Electrical Engineering} \\
\textit{National Taiwan University}\\
Taipei, Taiwan \\
hywei@ntu.edu.tw}
}

\maketitle

\begin{abstract}
An excellent self-driving car is expected to take its passengers safely and efficiently from one place to another. However, different ways of defining safety and efficiency may significantly affect the conclusion we make. In this paper, we give formal definitions to the safe state of a road and safe state of a vehicle using the syntax of linear temporal logic (LTL). We then propose the concept of safe driving throughput (SDT) and safe driving capacity (SDC) which measure the amount of vehicles in the safe state on a road. We analyze how SDT is affected by different factors. We show the analytic difference of SDC between the road with perception-based vehicles (PBV) and the road with cooperative-based vehicles (CBV). We claim that through proper design, the SDC of the road filled with PBVs will be upper-bounded by the SDC of the road filled with CBVs.

\end{abstract}

\IEEEpeerreviewmaketitle
\section{Introduction}\label{chap:111}
\begin{figure}[htbp]
	\begin{center}		\includegraphics[width=\columnwidth, height=3.0cm]{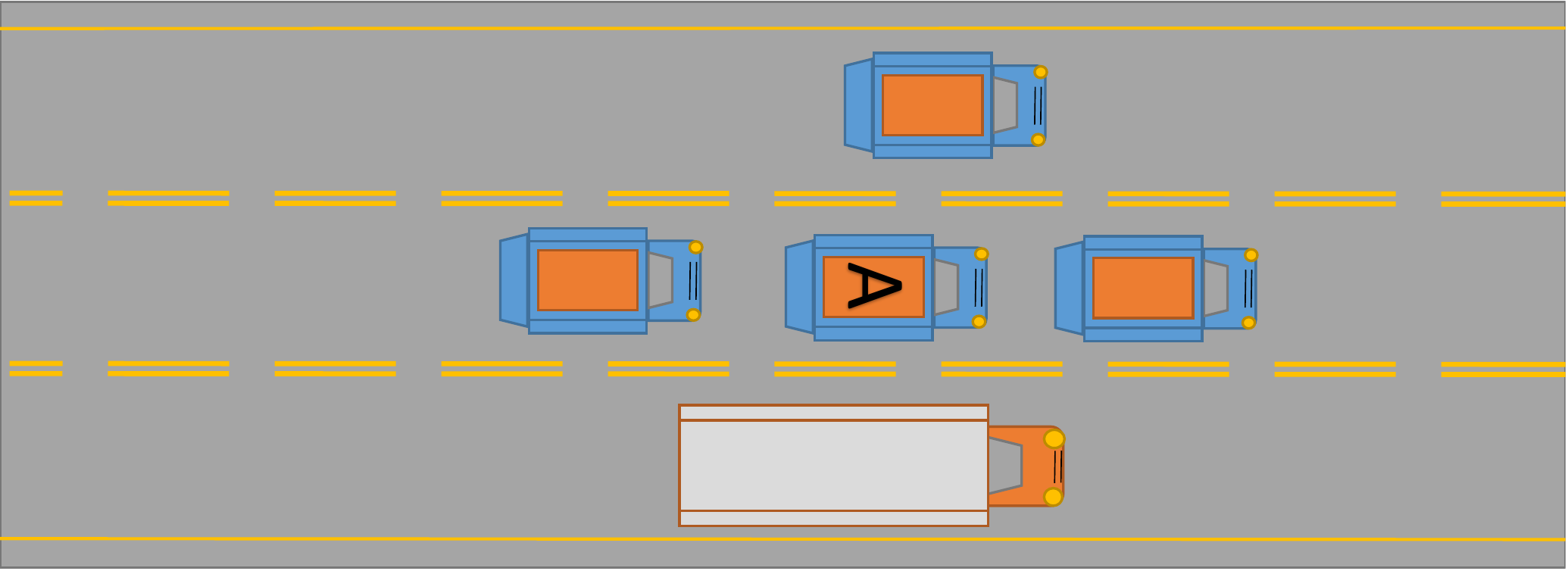}
		\caption{Absolute safety is not possible}
		\label{fig:1}
	\end{center}
\end{figure}

Self-driving car has been regarded as the solution to current transportation problems and has obtained significant improvement in recent years. Nevertheless, from the recent self-driving car accidents, people understand that safety remains an issue. From car A shown in Figure. \ref{fig:1}, it is clear that safety in the sense of collision cannot be guaranteed. The way to circumvent this problem is to redefine the meaning of safe. Shalev-Shwartz et al. define the safe in the sense of whether to share responsibility in an accident \citep{2017arXiv170806374S}. We adopt this idea of safe and create rules and definitions with mathematical rigors using LTL, the logic syntax often used in fields like automaton and control system to describe their concurrent characteristics with greater precision \citep{papa,Rabinovich2015,7039527}. 

Based on our definitions, we propose the concept of safe driving throughput (SDT) and safe driving capacity (SDC). SDT and SDC have more practical uses than the throughput defined conventionally since it guarantees the vehicles being calculated to be blame-free under a lowest speed limit. We analyze various factors that could have impacts on them by studying the longitudinal distance between vehicles. There were several related studies concerning the effects of longitudinal distance; \citep{4790743} focuses on how it affects the stability of platoons and \citep{article} focuses on passenger comfort affected by it. There were also several studies about rear-end collisions; \citep{SEPPELT20152621} focuses on the role the response of agents play and \citep{BELLA2011676} studies the relation between the warning system and the collisions. We aim to study how longitudinal distance affects rear-end collisions under the safe driving presumptions. 

\smallskip

The contributions of this paper are twofold. 
\begin{enumerate}
\item We formalize the concept of safe in the sense of responsibility using the LTL syntax. Based on it, we propose the concept of SDT and SDC that take both efficiency and safety into consideration. 
\item We show the fundamental differences between PBV and CBV by analyzing their SDC and present a protocol that could achieve such capacity.
\end{enumerate}

We organize the paper as follows. Section \ref{chap2} describes assumptions used in this paper and gives definitions needed in the following contexts, followed by our proposed protocol and detailed analysis on the SDC gain it brings to the road comparing to the case which the road is filled with PBVs in Section \ref{chap3}. In Section \ref{chap4}, we show the analytic result by figures and discuss them. In Section \ref{chap5}, we conclude our work and point out related open issues.

\section{Assumptions and Definitions}\label{chap2}
\subsection{\textbf{Assumptions}}
\subsubsection{\textbf{Vehicle Requirements}}\label{req}

All the vehicles we discuss are equipped with a high precision mapping system, navigation system, full autonomous controller, complete perception system, wireless interfaces for communication and high accuracy positioning system. The perception system consists of different types of sensors such as LIDAR, camera, and radar in charge of sensing the parameters of the car in front \citep{ars-3-205-2005}. The wireless communication interface allows the vehicle to use any of the wireless technology including cellular interfaces like 4G/5G and DSRC \citep{6515060,Xu:2004:VSM:1023875.1023879}. 

\subsubsection{\textbf{Homogeneous}}

All the vehicles are autonomous vehicles equipped with requirements mentioned above.

\subsubsection{\textbf{Reliable}}

All the messages from other vehicles are presumed to be reliable. 

\subsubsection{\textbf{Road Requirements}}

All the vehicles are running on a straight road without any intersection and merging point. 

\subsection{\textbf{Formalization and Definitions}}

In this subsection, we give definitions needed to define the safe driving throughput and safe driving capacity. For mathematical rigors, the syntax of LTL is used to help formalizing some of the definitions. In the following context, we assume that there are N vehicles $\omega_{1}, \omega_{2}, ... \omega_{N}$ on the road $R$, that is, $\Omega(R)=N$. Every vehicle $\omega_{i}$ has its own finite sequence of states $\sigma_{i} = <s_{i1},s_{i2},...,s_{iT}>$. The subscript $T$ denotes the time horizon we care about and it should be system-dependent. The predicate $C(\omega_{i})$ is $true$ iff $\omega_{i}$ collides with any other vehicle $\omega_{j|j\neq i}$ and the predicate $\Upsilon(\omega_{i})$ is true iff $\omega_{i}$ shares responsibility in an accident if any happens.

\bigskip
\noindent
\noindent \textbf{\textit{Definition 1: Longitudinal Distance:}} 

\noindent \textit{The longitudinal distance of two cars is the distance between their body center measured along the direction of the road.} 

\smallskip
\noindent

\noindent \textbf{\textit{Definition 2: best effort reaction(BER): }}

\noindent \textit{The best effort reaction (BER) of a car is to apply max braking power along the direction of the road until the car halts.}
\smallskip

\noindent \textbf{\textit{Definition 3: Safe Longitudinal Distance:}}

\noindent \textit{Two vehicles are in safe longitudinal distance if $\omega_f$, the one in front, makes a sudden change of behavior, the longitudinal distance between them is still sufficient for $\omega_r$, the one behind, to react and not bump into $\omega_f$. If even applying BER cannot prevent $\omega_r$ from the collision, such distance is unsafe.}
\smallskip
\noindent 

\noindent \textbf{\textit{Definition 4: Safe State of a Vehicle $\omega$:}}

\noindent\textit{A vehicle $\omega$ is in a safe state iff performing BER could spare it from responsibility even when an accident happens.}\\
\begin{equation}\begin{split}
\label{eq-1}
&\forall i\in \Omega(R); \ \ \omega_{i}\ is \ in\ the \ safe\ state \\ 
&\Leftrightarrow (\sigma_{i},t)\models\mathcal{G}_{[0,T]}(BER\Rightarrow( (C(\omega_{i})\Rightarrow\neg\Upsilon(\omega_{i}))\vee \neg C(\omega_{i}))\\
&\equiv (\sigma_{i},t)\models\mathcal{G}_{[0,T]}(BER\ \Rightarrow\neg\Upsilon(\omega_{i}))\\
\end{split}\end{equation}

\noindent \textbf{\textit{Definition 5: Safe State of a Road R:}}\\
\noindent\textit{A road R with N vehicles is in the safe state iff all the vehicles running on it are in safe state. }
\begin{equation}
\begin{split}
\label{eq-2}
&R \in safe\ state  \Leftrightarrow \forall i \in \Omega(R),\ \omega_i\ is \ in\ the \ safe\ state \\
& \equiv \forall \omega_{i}\in R,\ (\sigma_{i},t)\models\neg\mathcal{F}_{[0,T]}(BER\Rightarrow\Upsilon(\omega_{i}))   
\end{split}
\end{equation}

\noindent \textbf{\textit{Definition 6: Safe Driving Throughput (SDT) of a road R: }}\\
\noindent \textit{The safe driving throughput of R is the number of vehicles in the safe state that are on R. Noted that SDT(R) $\leq \Omega(R)$ and if the road $R$ is in safe state $\Rightarrow\ SDT(R)=\Omega(R)$.}

\noindent 
\smallskip
\noindent \textbf{\textit{Definition 7: Conservative Observation:}\label{def6}}

\noindent\textit{One observation is more conservative than another if the decision made based on it makes the vehicle more probable to stay in the safe state. We define the function $\Lambda(M)$ which takes observation metrics set M as input and return the most conservative one among the set as output.}
\begin{equation}
\begin{split}
\label{eq-3}
 \Lambda(M) & =\\
&  M_{ensemble} + \mathrm{argmin}{\textbf{\{B\}}}
\textit{$,\ \ if\  M \in \{V_f\}$}\\
&  M_{ensemble} + \mathrm{argmax}{\textbf{\{B\}}}
\textit{$,\ \ if\ M \in \{a_{\mathrm{max}brake}, L, \tau\}$}
\end{split}
\end{equation}

\noindent\textit{Here $M_{ensemble}$ is the ensemble average of all the observations and \textbf{B} is the set of all the biases of perception system.}

\smallskip
\noindent 
\noindent \textbf{\textit{Definition 8: Inaccuracy of Metrics:}}\label{def5}
\noindent\textit{Inaccuracy of metrics $M$ due to the perception system is defined as:}
\begin{gather}
\label{eq-1}
\Phi(M):=\frac{\left|\Delta M\right|}{\Lambda(M)}
\end{gather}
\noindent\textit{So that we have:}
\begin{gather}\label{eq0}
\Phi(M)=\frac{\left|M_{actual}-\Lambda(M)\right|}{\Lambda(M)}  
=\left|1-\frac{M_{actual}}{\Lambda(M)}\right| =\left|1-e_M\right|
\end{gather}
\noindent\textit{As shown above, the deviation of a metrics M, $e_M$, is defined as the ratio of its actual value and its most conservative estimate. We expect a good perception system to have $\Phi(M)\leq 5\%$, i.e., $0.95 \leq e_M \leq 1.05$.}

\smallskip
\noindent 
\noindent \textbf{\textit{Definition 9: Safe Driving Capacity (SDC):}\label{def7}}\\
\noindent\textit{We define the safe driving capacity of a road as the number
of vehicles on an M-kilometer-N-lane-straight-road requiring every vehicle to be in the safe state and runs at least V km/hr. We use the default values M=10, N=2, and V=100 unless otherwise specified and denote it as SDC(10,2,100).}

\smallskip
\noindent 
\noindent \textbf{\textit{Definition 10: Perception-based vehicle (PBV): }}\\
\noindent\textit{A perception-based vehicle is a vehicle that makes decisions based only on the data obtained from its perception system.}

\smallskip
\noindent
\noindent \textbf{\textit{Definition 11: Cooperative-based vehicle (CBV): }}\\
\noindent\textit{A cooperative-based vehicle is a vehicle that makes decisions mainly based on the data obtained through inter-vehicular communication. Fig. 2 shows the difference between roads with PBVs in definition 10 and road with CBVs in definition 11.}
\noindent
\begin{figure}[htbp]
	\begin{center}
		\includegraphics[width=\columnwidth, height=3.4cm]{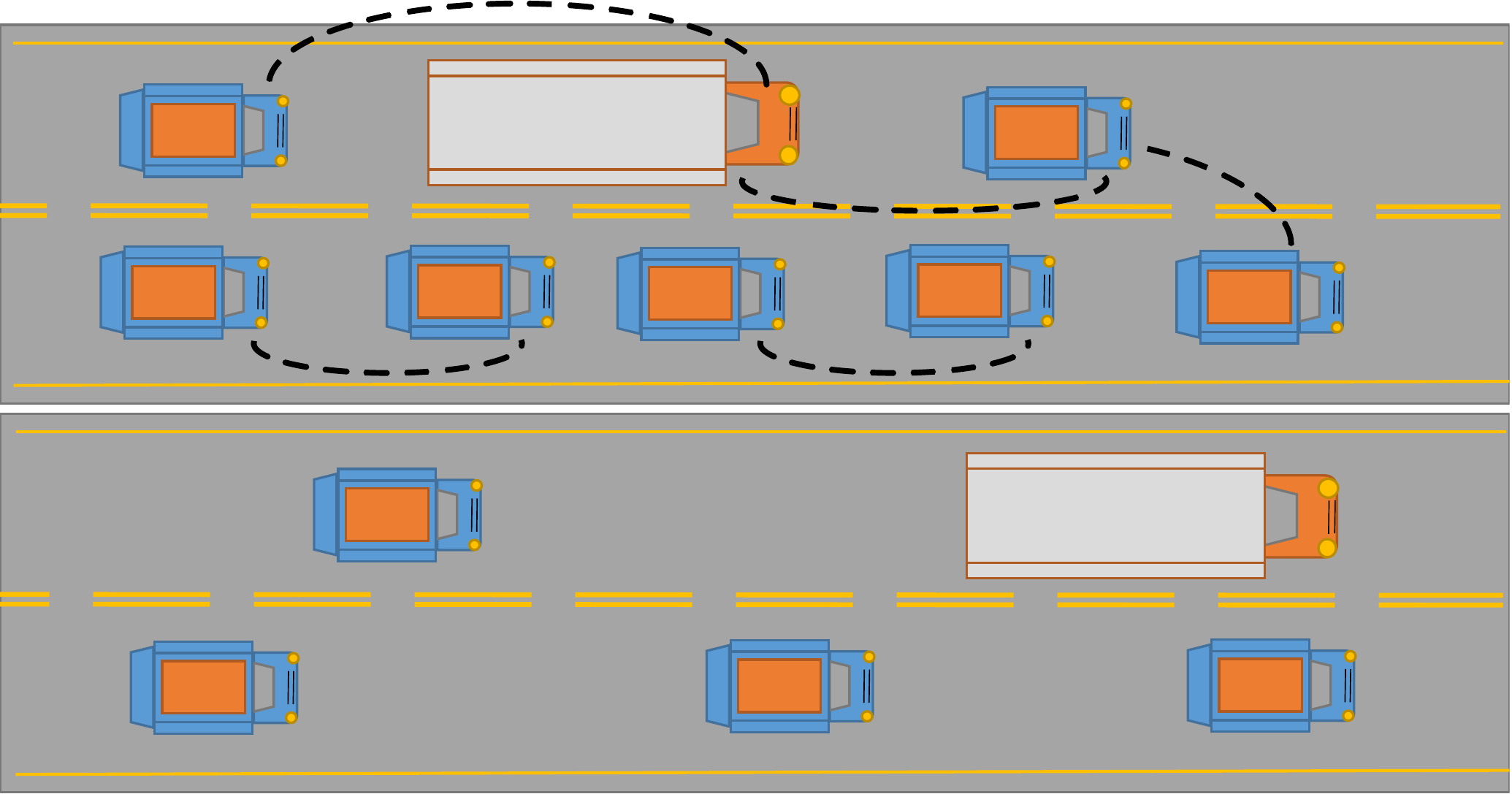}
		\caption{Traffic with and without inter-vehicle communication.}
		\label{fig:2}
	\end{center}
\end{figure}

\section{Derivations and Cooperative protocol} \label{chap3}
\subsection{\textbf{SDC analysis on road with PBVs}}\label{sec:4_1}

Based on the definitions in Section \ref{chap2}, we derive the formula of safe longitudinal distance between two vehicles with the variables given in TABLE \ref{tab2}. For simplicity, We define:  
\begin{equation} \label{eq1}
T_r=\tau +\frac{V_{r\_\tau(\eta) \_max}}{a_{maxbrake}}
\end{equation}

\begin{equation} \label{eq2}
T_f=\frac{V_f}{a_{maxbrake}}
\end{equation}

\begin{equation} \label{eq3}
V_{r\_\tau(\eta) \_max}=V_r+\tau(\eta) a_{maxacc}
\end{equation}

Here $T_r$ stands for the time elapsed from the moment it detects a sudden full brake from the car in front till the time $V_r=0$. $T_f$ is the time for the front car to enter full stop from its original speed, i.e. the time $V_f\to 0$. $V_{r\_\tau \_max}$ is the maximum velocity the rear car could after the entire response time $\tau$. With these derived variables, the safe longitudinal distance is given as: 
\begin{equation} \label{eq4}
\begin{split}
if\ &\ T_f<T_r:\ \\
& D_{Longitudinal-safe} = L + \frac{1}{2}(V_r+V_{r\_\tau(\eta) \_max})\tau(\eta)\\
& + \frac{1}{2}(T_r - \tau(\eta))V_{r\_\tau(\eta) \_max} -\frac{1}{2}V_f T_f \\ \\
if\ &\ T_f\ge T_r: \\
& D_{Longitudinal-safe}=\ L 
\end{split}
\end{equation}
\noindent \textbf{\textit{Proof of equation.\ref{eq4}:}}
For\textit{ }$T_f\ge T_r$, the proof is trivial if we presume both their speed drop at constant rate $a_{maxbrake}$ since it'll take longer for the front car to halt.

For $T_f<T_r$, Let $d_{T_r}$ denote the distance between two vehicles at time $T_r$.
Then 
\begin{equation} \label{eq6}
\begin{split}
d_{T_r} & = {\frac{1}{2}\tau(\eta) }^2a_{maxbrake}+ \frac{1}{2}\tau(\eta)(V_{r\_\tau(\eta) \_max}-V_r)\\
& + d_0+T_f\left[\left(v_f-v_{r\_\tau(\eta) \_max}\right)-\tau(\eta) a_{maxbrake}\right]\\
& -\frac{1}{2}\left(T_r-T_f\right)\left(v_{r\_\tau(\eta) \_max}-\left(T_f-\tau(\eta) \right)a_{maxbrake}\right)
\end{split}
\end{equation}

\noindent \textbf{\textit{}}

\noindent 

As long as two cars are still moving after $t$ seconds where $t>\tau(\eta) $, the first four terms on the right-hand side of equation \ref{eq6} will be $d_{Tf}$, meaning the distance between two vehicles after $T_f$. The last term in equation \ref{eq6} means the distance moved by the rear car from time $T_f$ to $T_r$. Now it only requires $d_{T_r}>L$, we can make sure two cars won't collide. By rearranging the terms, we conclude the proof. 

\subsection{ Cooperative protocol}\label{sec:4_2}
In this subsection, we propose a protocol that allows the road with CBVs to achieve its maximum SDT, i.e., SDC. 

\begin{algorithm}
\caption{Cooperative Protocol}
\label{alg1}
\begin{algorithmic}
\While{The car is still driving}
\State Send request to the front car $\omega_f$ for Information
\If{Receive response}
\State{Adjust distance according to the response}
\Else 
\If{No response but perception system works}
\State{Adjust distance based on $\Lambda(M)$}
\State{$where\ M \in perception\ observations$}
\Else 
\State{Adjust distance according to the most }
\State{conservative arguments predefined}
\EndIf
\EndIf
\EndWhile
\end{algorithmic}
\end{algorithm}

\subsection{SDC analysis on road with CBVs running Algorithm \ref{alg1}}\label{sec:4_3}

The accuracy of $V_f$, $a_{maxbrake}\ ,L$ and $\tau(\eta)$ could be enhanced by the cooperative inter-vehicle communication. And since the delay is additive, we have  
\begin{equation} \label{eq20}
\tau(\eta) = \tau_0 + \eta 
\end{equation}

Here $\tau_0$ denotes the time from the point $\omega_r$ receives information from $\omega_f$ till its system starts to brake. 
To show the clear contrast to the result from road with PBVs, we denote all the variables with subscript C as the actual value corrected by communication. By definition of inaccuracy of metrics defined in Section \ref{chap2}, we have:
\begin{equation} \label{eq8}e_L = 1 - \Phi(L) = L_C/L\end{equation}
\begin{equation} \label{eq9}e_{V_f} = 1 + \Phi(V_f) = V_{fC}/V_f\end{equation}
\begin{equation} \label{eq10}e_{brake} = 1 - \Phi(brake) = a_{maxbrakeC}/a_{maxbrake}\end{equation}
\begin{equation} \label{eq11}e_\tau = 1 - \Phi(\tau) = {\tau_C}/\tau_0\end{equation}

\noindent Noted that if the deviations $e_L$, $e_{V_f}$, $e_{brake}$ and$\ e_\tau$ are all 1, it means the PBVs perceive the actual value of all the vehicle arguments needed. Also noted that $\eta$ is not affected by $\tau_C$. By the same procedure in subsection A: $ V_{r\_\tau \_max\_C}=Vr+(e_{\tau }\tau_0+\eta) a_{r\_\mathrm{max}\mathrm{}\_acc} $ ; 
$ T_{rC}=e_{\tau }\tau_0+\eta+{V_{r\_\tau \_max\_C}}/{a_{maxbrake}}$, and $ T_{fC}={V_{fC}}/{a_{maxbrakeC}}$. Using these variables, the communication-corrected version of safe longitudinal distance $D_{corrected}$ will be $(L+e_L)/2$ for the trivial case $T_f\ge T_r$ ; for the case $T_f < T_r$,

\begin{equation} \label{eq19}
\begin{split}
& D_{corrected} = \frac{1}{2}\{(L+Le_{L}) - (V_{fC} T_{fC})\\
& + (V_r + V_{r\_\tau \_max\_C})(e_\tau \tau_0 + \eta) + \frac{V_{r\_\tau \_max\_C}^2}{a_{maxbrake}} \} \\
\end{split}
\end{equation}

\newtheorem{theorem}{Theorem}
\newtheorem{lemma}{Lemma}
\noindent\begin{lemma}\label{lemma1}
If for any metrics $M$, $\Lambda(M)$ is always more conservative than its actual value, the estimation of the safe longitudinal distance of CBVs is less or equal to the one of PBVs:
\begin{equation*}
D_{corrected} \leq D_{Longitudinal\_safe}
\end{equation*}
\end{lemma}
\noindent \textbf{\textit{Proof of Lemma \ref{lemma1}:}}

It's the direct consequence of \textit{definition 5} and \textit{definition 6}.

\DeclarePairedDelimiter\floor{\lfloor}{\rfloor}
\noindent\begin{lemma}\label{lemma2}
Let $E\{X\}$ denote the expected value of variable X,
The SDC of a road R with PBVs is given as:
\begin{equation*}
\\ SDC(R(PBV)) = \floor{\frac{N(M-L)} {E_{\omega_i|\forall i \in \Omega(R)}\{D_{longitudinal\_safe}\}}}+1
\end{equation*}
\end{lemma}
\noindent\begin{lemma}\label{lemma3}
The SDC of a road R with CBVs is given as:
\begin{equation*}
\\ SDC(R(CBV)) = \floor{\frac{N(M-L_C)} {E_{\omega_i|\forall i \in \Omega(R)}\{D_{Corrected}\}}}+1
\end{equation*}
\end{lemma}

\noindent \textbf{\textit{Proof of Lemma \ref{lemma2} and Lemma \ref{lemma3}:}}
The $(M-L)$ term in the fractions denotes the distance from the center of the first vehicle to the center of the last vehicle. Since all the vehicles are asked to keep at least $D_{longitudinal\_safe}$ from the vehicles in their front, we have $\floor{\frac{(M-L)} {E_{\omega_i|\forall i \in \Omega(R)}\{D_{longitudinal\_safe}\}}}$ inter-vehicle spaces. This implies that a single lane can accommodate $\floor{\frac{(M-L)} {E_{\omega_i|\forall i \in \Omega(R)}\{D_{longitudinal\_safe}\}}}+1$ vehicles. 
Proof of Lemma \ref{lemma3} is identical to Lemma \ref{lemma2}.

\noindent\begin{theorem}\label{thereom1}
The $SDC(R(PBV))$ is upper bounded by the $SDC(R(CBV\ using\ our\ protocol))$ if $\Lambda(M)$ is always more conservative than actual $M$ for any metrics $M$.
\begin{equation*}
\\ SDC(R(PBV))\ \leq \ SDC(R(CBV)) 
\end{equation*}
\end{theorem}

\noindent \textbf{\textit{Proof of Theorem \ref{thereom1}:}}
Since $\Lambda(M)$ is always more conservative than actual $M$, $0\leq E\{L_C\} \leq E\{L\}$, and from Lemma \ref{lemma1}, we have $0\leq D_{corrected} \leq D_{Longitudinal\_safe}$. Based on these two inequalities, we complete the proof.

\section{Simulation and Discussion} \label{chap4}
\begin{figure*}[htbp]
    \begin{center}
  \makebox[\textwidth]{\includegraphics[width=\textwidth, height=6.0cm]{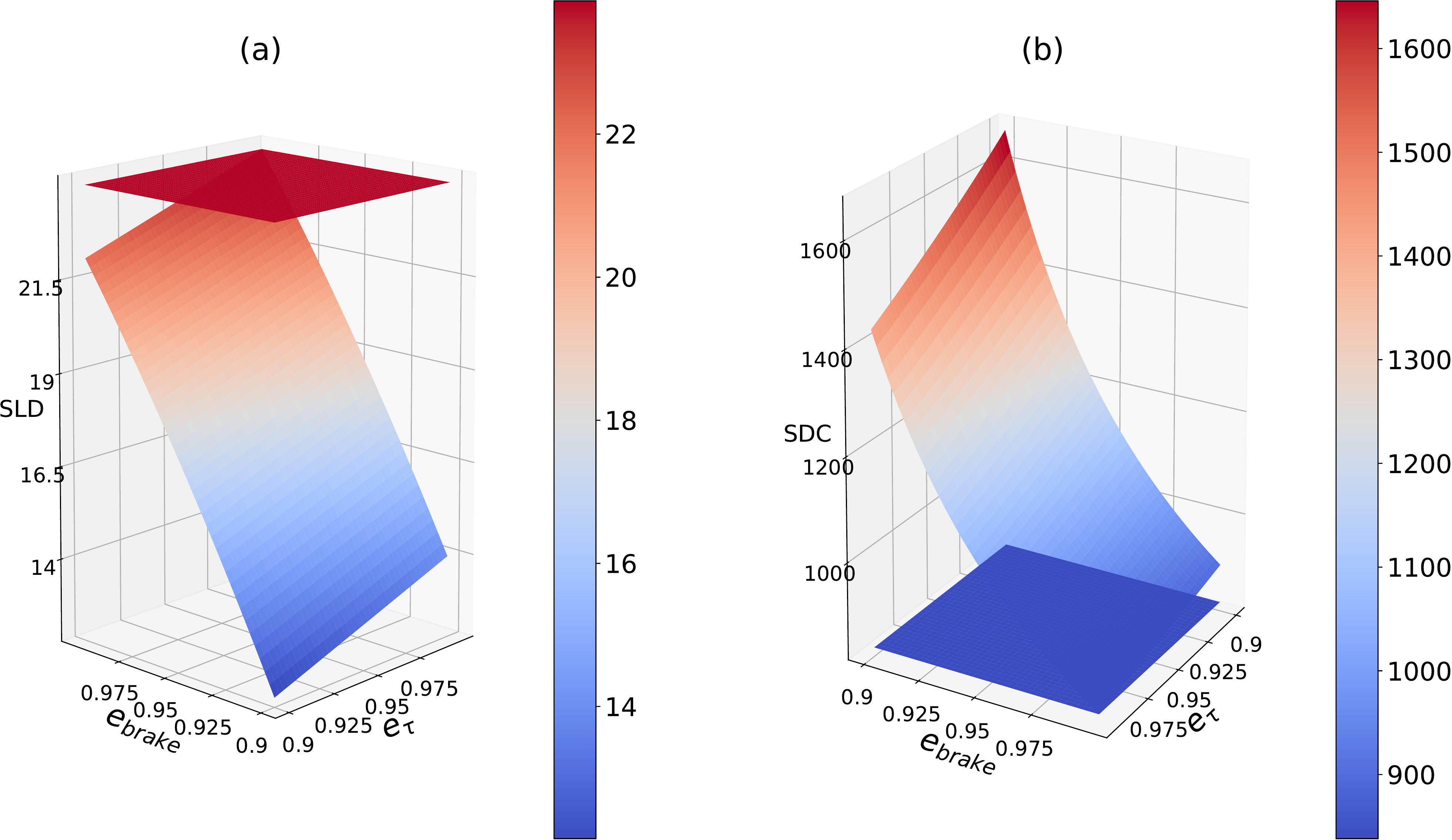}}
    \caption{Situation when $e_V = 1$. In both (a) and (b), the plane parallel to the ground is the result of PBV and the colorful hyper-plane is the result of CBV. (a) shows the minimum safe longitudinal distance and (b) shows the corresponding SDC.}
  \label{fig:eT1}
\end{center}
\end{figure*}

\begin{figure*}[htbp]
    \begin{center}
  \makebox[\textwidth]{\includegraphics[width=\textwidth, height=6.4cm]{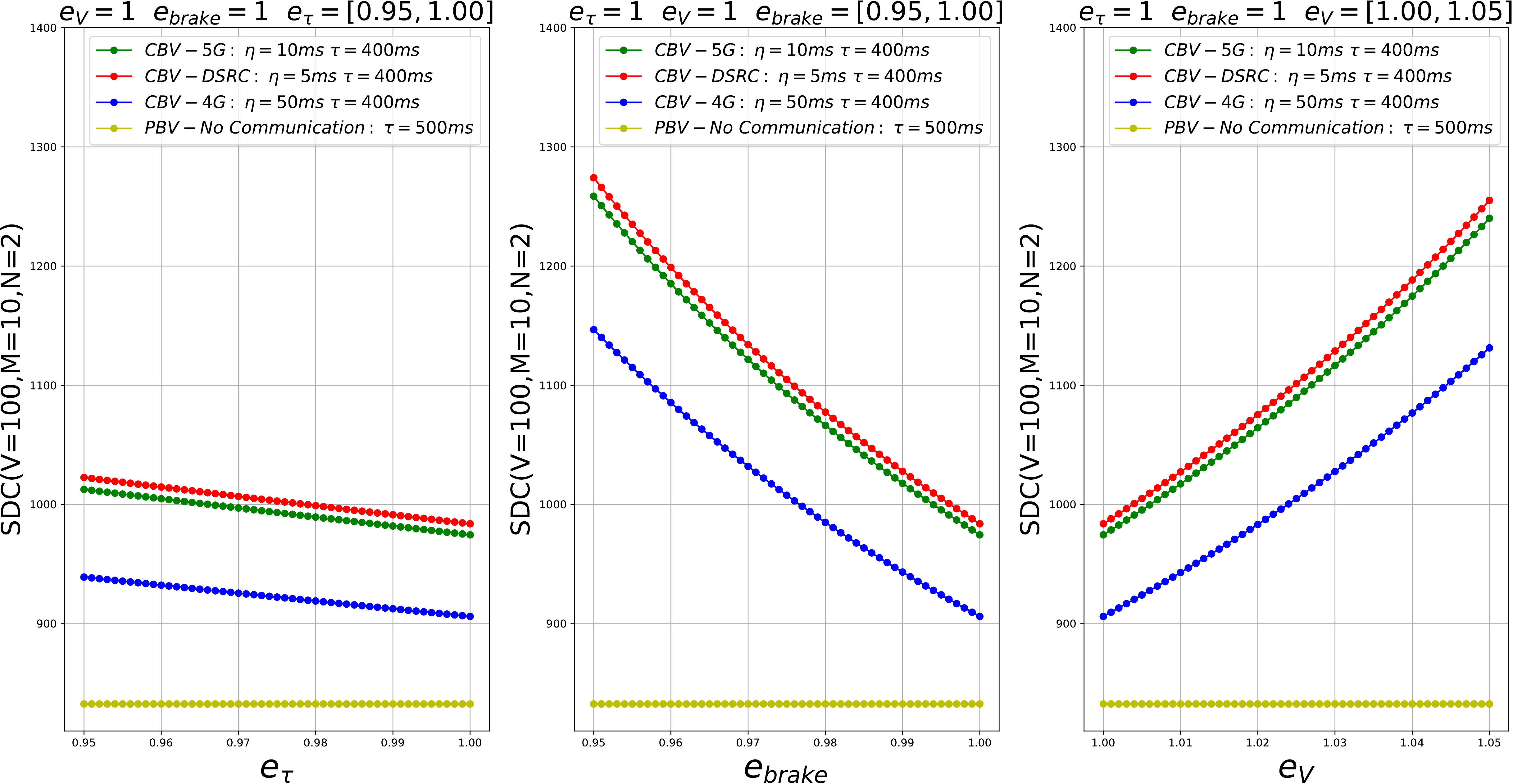}}
    \caption{Comparison of different kinds of communication schemes and different observation inaccuracies.}
  \label{fig:eT2}
\end{center}
\end{figure*}

For a vehicle equipped with ABS (Anti-lock brake system), at the speed of 100 km/h, the maximum acceleration and deceleration are $2.2\ to\ 4.0\ m/s^2$ and around $9\ m/s^2$ respectively \citep{6576926}. In our analytic simulation, we let $\tau_0 = 0.5$ sec for the road with PBVs and $\tau_0 = 0.4$ for the road with CBVs. we evaluate the SDC(10,2,100). Noted that if the lowest speed limit V is not put in the definition, the SDC becomes meaningless since the capacity achieve maximum if all the vehicles stop. From Fig. \ref{fig:eT1}, we can see that even some minute perception inaccuracies could lead to huge differences in the safe longitudinal distance and thus SDC. Fig. \ref{fig:eT2} shows the effect on SDC from various values of $\eta$ under different inaccuracies of $e_\tau$, $e_{brake}$ and $e_V$. Each $\eta$ here features a specific kind of V2V communication. The latency $\eta$ of DSRC and 5G in V2V applications are supposed to be shorter than the value we adopt \citep{7876982}. The machine response time $\tau_0$ we use here lies in the range of $[400, 500]$ ms. This is the around $\frac{1}{3}$ the time needed for an inattentive driver and $\frac{1}{2}$ of an attentive driver \citep{doi:10.1518/001872001775898250}. If we compare Figure. \ref{fig:eT1} with Figure. \ref{fig:eT2}, we can conclude that even some inaccuracy metrics have little impact on the SDC, Their effects become ineligible when all of them are considered.

\section{Open issues and future work}\label{chap5}
In this work, we intentionally simplify some of the analysis, especially the parts that are supposed to be probabilistic instead of deterministic due to the stochastic nature of both en-route drivings and inter-vehicle communications. The scenarios we presumed in Section \ref{chap2} might also be too ideal from the perspective of the real-world situation. The SDC we obtained is the one without any concept of platooning but focus on the result of individual car behaviors. Training a self-driving car is itself a challenging problem but we choose to ignore this issue and presume the vehicles are capable of driving perfectly like an adrift human driver. For the convenience of deterministic analysis, we let the road to be intersection-free and straight. However, the case with intersections and roads that are not straight might generalize the SDT and the SDC. We leave them as our future work. Another thing worth mentioning is that the concept of SDT and SDC may also apply to human drivers or the heterogeneous situation that consists of man-drive vehicles and self-driving vehicles with just minor modifications. This fact makes our work even more practical.   

\begin{table}[htbp]
\caption{Variables in Section \ref{chap3}}
\begin{center}
\begin{tabular}{|c|c|}\hline
{\bf Variable (unit)} & {\bf Description} \\ \hline
{\bf $(\sigma,i)\models \phi$} & $\phi(\sigma(i))$ is true \\ \hline
{\bf $(\sigma,i)\models \mathcal{G}_{[a,b]}\phi$} & $\forall j\in [i+a,i+b], (\sigma,j)\models \phi$ \\ \hline
{\bf $(\sigma,i)\models \mathcal{F}_{[a,b]}\phi$} & $\exists j\in [i+a,i+b], (\sigma,j)\models \phi$ \\ \hline
{\bf $(\sigma,i)\models\phi \wedge \psi$} & $(\sigma,i)\models \phi \wedge (\sigma,i)\models \psi$\\ \hline
{\bf $(\sigma,i)\models\phi \vee \psi$} & $(\sigma,i)\models \phi \vee (\sigma,i)\models \psi$\\ \hline
{\bf $D_{Longitudinal-safe}\ (m)$} & The minimum safe longitudinal distance \\ \hline
{\bf $\tau(\eta)\ (sec)$} & Response time for rear car \\ \hline
{\bf $V_r\ (m/s)$} & Speed of rear car \\ \hline
{\bf $V_f\ (m/s)$} & Speed of front car \\ \hline
{\bf $a_{maxbrake}\ (m/s^2)$} & The deceleration of full braking power\\ \hline
{\bf $a_{maxacc}\ (m/s^2)$} & Max acceleration of a car\\ \hline
{\bf $L\ (m)$} & Length of vehicle, we suppose all the same. \\ \hline
{\bf $T_r\ (sec)$} & The time rear car need to enter a full stop \\ \hline
{\bf $T_f\ (sec)$} & The time front car need to enter a full stop \\ \hline
{\bf$V_{r\_\tau(\eta) \_max}\ (m/sec)$} & {Maximum possible $V_r$ after $\tau(\eta)$ sec}  \\ \hline
{\bf $D_{corrected}\ (m)$} & The corrected minimum safe longitudinal distance\\ \hline
{\bf $\tau_C\ (sec)$} & Corrected Response time\\ \hline
{\bf $V_{fC}\ (m/s)$} & Corrected Speed of front car \\ \hline
{\bf $e_L$} & Inaccuracy of sensors measuring L \\ \hline
{\bf $e_{V_f}$} & Inaccuracy of sensors measuring $V_f$ \\ \hline
{\bf $e_{brake}$} & Inaccuracy of sensors measuring $a_{maxbrake}$  \\ \hline
{\bf $e_{\tau C}$} & Inaccuracy of sensors measuring $\tau$  \\ \hline
{\bf $a_{maxbrakeC}\ (m/s^2)$} & Corrected max deceleration of front car\\ \hline
{\bf $L_C(m)$} & Corrected length of vehicle \\ \hline
{\bf $T_{rC}\ (sec)$} & Corrected time rear car need to enter a full stop \\ \hline
{\bf $T_{fC}\ (sec)$} & Corrected time front car need to enter a full stop \\ \hline
{\bf$V_{r\_\tau \_maxC}\ (m/s)$} & {Corrected maximum possible $V_r$ after $\tau_C+\eta$ sec}  \\ \hline
{\bf $\eta\ (sec)$} & Random variable of the latency between CBVs \\ \hline

\end{tabular}
\label{tab2}
\end{center}
\end{table}
\noindent

\bibliographystyle{IEEEtran}
\bibliography{ref}

\end{document}